# $R$-band imaging of fields around $1 < z < 2$ radiogalaxies [1]

N. Benítez, E. Martínez-González & J.I. González- Serrano

Dpto. de Física Moderna, Universidad de Cantabria and

Instituto de Física de Cantabria, CSIC-Universidad de Cantabria

Facultad de Ciencias, Avda. Los Castros s/n, 39005 Santander Spain

L. Cayón

Lawrence Berkeley Laboratory

1 Cyclotron Road, Berkeley, CA 94720, USA

---

[1]Based on observations made with the Isaac Newton Telescope

# ABSTRACT

We have taken deep $R$-band images of fields around five radiogalaxies: 0956+47, 1217+36, 3C256, 3C324 and 3C294 with $1 < z < 2$. 0956+47 is found to show a double nucleus. Our data on 1217+36 suggest the revision of its classification as a radiogalaxy. We found a statistically significant excess of bright ($19.5 < R < 22$) galaxies on scales of 2 arcmin around the radiogalaxies (which have $R \approx 21.4$) in our sample. The excess has been determined empirically to be at $\gtrsim 99.5\%$ level. It is remarkable that this excess is not present for $22 < R < 23.75$ galaxies within the same area, suggesting that the excess is not physically associated to the galaxies but due to intervening groups and then related to gravitational lensing.

*Subject headings:* Radiogalaxies, clustering, gravitational lensing

## 1. Introduction

The study of the environment of radiogalaxies has shown that these objects tend to lie in regions with density richer than average (Lilly & Prestage 1987). This enhancement is more pronunciated for FRI types, which often can be found in the cores of rich clusters, than for FRII sources. The situation changes at $z \sim 0.5$ (Hill & Lilly 1991, Yates et al. 1989). These authors have shown that both FRI and FRII sources lie in similarly rich environments at these redshifts, as it is also the case for radio-loud QSOs (Yee & Green 1987). On the contrary, radio-quiet QSOs reside in much poorer environments (Ellingson et al. 1991). This suggests that the cluster environment of objects at this redshift is strongly dependent on radio luminosity. Recent observations of both types of QSOs (Boyle & Couch 1993,Hintzen et al. 1991) have shown this characteristic to be also present at $0.9 < z < 1.5$. The environments of radio-selected BL Lac objects have been studied by Fried et al. (1993). They found clustering around objects at redshifts $z \lesssim 0.7$ but not for the $z \approx 0.9$ ones.

There have been several reports of statistically significant associations of high redshift QSOs and radiogalaxies with foreground objects: galaxies (3C324 is one of the best cases known Hammer & Le Fevre 1990, Thomas et al. 1994) Zwicky clusters (Seitz & Schneider 1994), IRAS galaxies (Bartelmann & Schneider 1993), X-ray photons taken from the ROSAT All Sky Survey (Bartelmann et al. 1994). These associations are interpreted as a lensing effect with the lens being either a single galaxy or clusters of galaxies. Powerful radio sources are particularly well suited to detect associations due to gravitational lensing, because of the steep slope of their number counts (Wu & Hammer 1994).

Here we search for excesses of objects around $1 < z < 2$ radiogalaxies with apparent magnitudes ($R \approx 21.4$) and spanning a range of one order of magnitude in radio luminosity. The paper is structured as follows: In Sec. 2. we describe the observations and data reduction methods. In Sec. 3. we give our number counts data. Sec. 4. reports the most remarkable characteristics of each radiogalaxy. Sec. 5. describes our main results about the excess around the radiogalaxies. In Sec. 6. we compare our observations with those of other authors. Sec. 7. summarizes our main results and conclusions.

## 2. Observations and data processing

Observations were carried out during the night of 1992 April 1 at the prime focus of the 2.5m Isaac Newton Telescope (INT) on the island of La Palma (Canary Islands, Spain). The detector used was an EEV $1280 \times 1180$ CCD camera with a scale of 0.57 arcsec/pixel and readout noise of 6 electrons. We took 7 exposures of 500 seconds each in the direction of the radiogalaxies 0956+47, 1217+36, 3C256, 3C294 and 3C324 using the Kitt Peak $R$-band filter. All the fields have $b \gtrsim 50°$. Photometric standard stars in NGC4147 were also observed in order to calibrate the frames. The night was not photometric and the seeing was approximately 1.2 arcsec FWHM. There is no vignetting in the field.

The final images, obtained by coadding the individual frames for each object, were flat-fielded using sky and dome exposures. Their final sizes, after removing the borders, are $1200 \times 1090$ pixels ($11.4 \times 10.3 \text{arcmin}^2$). Due to some scattered light the sky did not remain flat enough. In order to subtract the sky background we took a grid of $50 \times 50$ points evenly distributed over the images. We then calculated the median value in the frames in boxes of $13.7 \times 12.8$ arcsec$^2$ centered at those grid points and assigned that value to this central point. A bi-cubic spline algorithm was used to construct a sky frame which was finally subtracted from the frames. Cosmic rays were removed automatically.

We searched for objects in the frames using the standard package PISA (Position Intensity and Shape Analysis, Draper & Eaton 1992). Our detection threshold was of 9 connected pixels above $1\sigma$ level per pixel. Objects close to the borders and within areas contaminated by the break-up of bright stars and satellite traces have not been considered.

We also have removed an area $\approx 10\%$ of the frame surface on the upper right corner of all the fields due to scattered light. In order to test our detection algorithms we generated frames with Poisson noise. From this simulations we estimate the number of spurious detections as $\lesssim 7\%$. This estimation was also confirmed by the results of the search for 'negative' objects in the real images.

## 3. Number counts

The sum of the number-counts in the five fields (Fig. 1.) versus $R$-magnitude has a slope of $\approx 0.37$ (within the magnitude range $20.5 < R < 23.75$) and follows very well the results of other authors (Metcalfe et al. 91). Our adopted completeness limit is $R = 23.75$. Due to changing photometric conditions along the night our error in magnitude calibration from field to field is $\lesssim 0.2$ mag. The errors plotted in Fig. 1 are the rms of the background counts from field to field within each magnitude bin. This variance is consistent with our assumed photometric error. It is noteworthy that fortunately, this calibration error does not affect our main result about the excess of objects around the radiogalaxies; as we would explain later in section 5, for a given radiogalaxy we determine the expected background counts using only the field of that radiogalaxy. This means that assuming that the photometry is coherent within each field (no vignetting is found), our only problem is the inaccuracy in the limits of the magnitude ranges.

The radiogalaxies have magnitudes $21.2 < R < 21.6$. We did not perform star-galaxy separation, but excluded from the following analysis all objects with $R < 19.5$. The amount of stars in the remaining data is only a few percents of the total number counts.

## 4. Notes on individual objects

As we have mentioned in the Introduction, it is not clear whether the galaxies clustering in projection around AGN are physically associated with them. This poses a problem: which are the optimal magnitude range and angular scale to search for these excesses? It is evident that they cannot be the same for near, foreground objects than for a $z > 1$ ones. Another problem is that using a fixed window we lose information about the individual characteristics of the excess around each radiogalaxy. On the other hand, in order to give the statistical significance of the excesses around our sample of radiogalaxies a fixed angular scale and magnitude range must be used. We therefore have combined two approaches: In Sec. 5. we have performed a statistical study of the general characteristics of the excess around our whole sample using two pre-chosen fixed angular scales and a magnitude range. In this section, on the contrary, we have deliberately varied the magnitude range and angular scale for each radiogalaxy in order to maximize the excess over an assumed Poisson distribution. This may create a galaxy excess, even if no excess galaxies are present and we therefore in no way try to attach any statistical significance to the excess results of this section; they are merely descriptive. We are just trying to find out whether there are any preferred scales and magnitude ranges for the presence of clustering and extract all the information our data can give us about the characteristics of the excess around each separate radiogalaxy.

And so, we shall characterize the apparent clustering in a given region and within a particular range of magnitudes comparing the number of objects found with their expectation $N_{\text{exp}}$. In order to find $N_{\text{exp}}$ we first calculate the 'unspoilt' surface within the considered region by laying 100,000 random points over the field and applying to them the same exclusion criteria as to the real objects. Afterwards we multiply this value by the average density of objects within the given range of magnitudes on that very field. A rough estimation of the rms value for the expected number counts $N$ within a given region is $1.5\sqrt{N}$.

### 4.1. 0956+47

An unresolved radiosource (Vigotti et al. 1989), this radiogalaxy was optically identified (McCarthy 1991) at a redshift of $z = 1.026$. Our observations(Fig. 2) show that $0956 + 47$ has a double structure formed by two components with maxima separated by 1.8 arcsec which are responsible for the apparently elongated shape mentioned by McCarthy. The object A of McCarthy's identification chart, at 24 arcsec from 0956+47 is in fact a disk galaxy ($R = 18.24$) with a very bright and apparently unresolved nucleus.

McCarthy already pointed out the existence of an apparent excess of faint objects around this radiosource. We found 7 objects against 2.5 expected with $21 < R < 23.75$ ($R_{0956+47} = 21.22$) within a radius of 15 arcsec around it. Clustering seems to be present — although not so clearly — at greater scales: within 165 arcsec there are 342 against 308.3 with $21 < R < 23.75$. At $19.5 < R < 21$ we find 36 instead of 28.2.

The double structure of 0956+47, the appearance of the nearby objects and the fact that they seem to be fainter than 0956+47 strongly suggest that we probably are observing a group of galaxies at $z = 1.026$.

### 4.2. 1217+36

This object has been recently resolved at 15 MHz (Naundorf et al. 1992). 1217+36 is classified as a core-jet source or, more unlikely, a double unresolved. Optically, this object was first identified by Allington-Smith et al. (1982). They measured a magnitude $m_r = 22.3 \pm 0.7$ in the Wade system and an angular diameter of 3? arcsec (they clearly had some reserves as the question mark shows). The object was classified as G?. Our observations give a magnitude $R = 21.6$, what corresponds well with the transformation $m_r - R \sim 0.8$ for high redshift radiogalaxies given by Allington-Smith et al., and a FWHM =

1.2 arcsec equal to the seeing size. The object thus seems to be stellar-like (Fig. 2b.), what casts some doubts on its classification as a radiogalaxy and therefore on the estimation of its redshift ($z = 1.2$, Lilly et al. 1985) on the basis of the $K - z$ relation. The determination of its type and redshift should be confirmed by spectroscopic observations.

Allington-Smith et al. (1982) found an excess of $m_r \lesssim 23$ objects at a 90% level in their whole field excluding the zone with radius $< 50$ arcsec around 1217+36 (their image is $3.6 \times 3.6$ arcmin$^2$). They estimated the expected background in this region as the mean from 47 such fields and found it to be 37.4 objects. Unfortunately they do not mention the exact number of counts within this region for the 1217+36 field, but assuming a Poisson statistics, from their level of significance it must be within the range 48-53. In order to compare our results with theirs, we counted objects in the same region within the approximately corresponding magnitude range $19.5 < R < 22.2$. We obtained a good agreement for the expectation from the number counts in our whole field, which is 38. However, the number of objects actually found is only 45. We also do not detect a strong excess in this region at fainter magnitudes: we found 145 objects with $22.2 < R < 23.75$ instead of 130.

Allington-Smith and colaborators did not detect any excess with $> 90\%$ significance within the central region $r < 50$ arcsec. This means that they found $< 13$ objects. On contrast, our image shows 15 objects within the range $19.5 < R < 22.2$ against an expectation of 7.9 in this region. This excess is maximized within the range $19.5 < R < 21.6$ and in the area with radius less than 60 arcsec: 18 objects where only 6 are expected. Such an excess is not present for fainter objects: for $21.6 < R < 23.75$ there are 48 objects instead of 42.

These differences between our work and that of Allington-Smith et al., which do not seem to involve a great amount of objects are not difficult to explain taking into account the different photometric systems used and the rather big calibration uncertainties present in both studies.

### 4.3. 3C256

This source is an asymmetric double with a radio angular size of 4.2 arcsec (McCarthy et al. 1991). Its identification is a galaxy at redshift $z = 1.819$ (Spinrad & Djorgovski 1984b). Its shape is remarkably elongated and formed by aligned multiple components, what opens the possibility of 3C256 being a gravitational lensed object (Le Fevre et al. 1988). Our measure of $R = 21.51$ is consistent with the value found by these authors: $R = 21.57 \pm 0.1$. We serendipitously found a cluster of faint galaxies $\approx 4$ arcmin from 3C256. The number of objects in the cluster considerably raises (by a 7%) the average density of the field and masks the excess around the radiogalaxy. For example, if we exclude an area $2.8 \times 3.8$ arcmin$^2$ centered on the cluster from our consideration, we find 205 objects with $21 < R < 23.75$ instead of 170 within 120 arcsec around 3C256. The expectation without excluding the cluster is 182 objects.

### 4.4. 3C294

3C294 is an asymmetrical double radiogalaxy (14.5 arcsec between lobes) with a weak core (McCarthy et al. 1990). It is identified with an 150 kpc Ly$\alpha$ cloud and a very red compact object in the $K$-band nearly coincident with the radio core. Its observation is complicated by the presence of a $V = 12$ mag star 10 arcsec from the radio centroid. The halo around this star due to the saturation filled more than a $1 \times 1$ arcmin$^2$ area in our image. After smoothing and substracting this halo the radiogalaxy is still not visible , but now the area affected by saturation is only $\approx 10\%$ of the $2 \times 2$ arcmin$^2$ box centered on the radiogalaxy.

In the remaining area there is a slight excess of $19.5 < R < 21$ objects within a radius of 45 arcsec around the assumed position of 3C294. We found 4 where 1.4 are expected.

### 4.5. 3C324

This asymmetrical double radiogalaxy, with a lobe-to-lobe size of 8 arcsec, was first identified by Kristian et al. (1974). It has a redshift $z = 1.206$ and a $V$-magnitude of 22.6 (Spinrad & Djorgovski 1984a). We measure $R = 21.4$. This object has a multiple structure, with its central part being formed by a galaxy with $z = 0.84$. This makes 3C324 one of best gravitational lens candidates found so far (Le Fevre et al. 1987, Hammer & Le Fevre 1990).

Since its identification, several authors have suggested that 3C324 is a central cluster galaxy. Kristian et al. (1974) mention a likely cluster with $N \approx 10$ in their $V$ plate although this cluster is not visible in their finding chart as Spinrad & Djorgovski (1985) noted. These authors apparently detected several objects that could be companion galaxies on the spectrum of 3C324 and obtained deep images of the 3C324 field which allowed them to label several objects as probable cluster members. Le Fevre et al. (1987) obtained deep broad-band images of 3C324 with the CFHT. Their exposure times were 4,500 s in $R$ and 5,400 s in $I$. They found more than 100 galaxies in $1.15 \times 1.86$ arcmin$^2$ around 3C324 and then affirm that 'this confirms the hypothesis of a cluster around 3C324'. However, they do not have any comparison field to obtain the expected background value, and thus, their conclusions about the existence of a cluster — as those of the above mentioned authors — are highly subjective.

Unfortunately these authors do not describe their detection algorithm nor explain in which band was performed the detection. Our $R$-band observations are not so deep as those of Le Fevre et al. and we therefore find much less objects than they do in the same region. PISA detects only 44 objects in the central $1.15 \times 1.86$ arcmin$^2$ around 3C324, which is far from being an excess: the expectation is 41.5. For $19.5 < R < 23.75$ and radius $< 45$ arcsec we find 27 objects where 25.8 are expected. Only if we look at magnitudes similar or brighter than the radiogalaxy we observe some excess: 11 objects with $19.5 < R < 22$ instead of 6.1, whereas for $22 < R < 23.75$ there are 16 objects where expected 19.7 (for $R > 22$ there are 20 objects detected instead of 29.8).

Further spectroscopic and multiband observations are needed, but the evidence seems to be against a cluster at redshift 1.2. Given that the scarce clustering apparently found tends to have $R \lesssim R_{3C324}$, and that $R_{3C324}$ is supposedly magnified by the gravitational lensing effect, it is very likely that if there is any real excess this is due to an intervening group, probably at the same redshift as the gravitational lens.

## 5. Projected clustering around $1 < z < 2$ radiogalaxies

We have studied the clustering of objects in 2 arcmin × 2 arcmin boxes around each radiogalaxy. This corresponds to a scale of $\approx 1$ Mpc ($\Omega_o = 1, h = 0.5$), at redshift $1 < z < 2$. We have divided the objects into two groups, one with magnitudes similar or brighter than the radiogalaxies: $19.5 < R < 22$, and another with fainter magnitudes: $22 < R < 23.75$. The data for the whole fields are listed in table 1. The main results for the 2 arcmin × 2 arcmin boxes are listed in table 2. In order to find the expected background value ($ne$ in Table 2) we multiply the 'unspoilt' surface within the considered central region by the average density of galaxies within the given range of magnitudes on that field. We determined the rms empirically, although many of the quoted authors find the statistical significance taking the variance as if the distribution of galaxies were poissonian, what certainly it is not the case. Galaxies cluster, and their angular two-point correlation function has the form: $\omega(\theta) = A\theta^{-\delta}$. The true variance of the galaxy distribution thus depends not only on the number of objects (as in the poissonian case) but also on $\omega(\theta)$ and on the value of the studied area $\Omega$. It is equal to

$$\left\langle \frac{N - \langle N \rangle}{N} \right\rangle^2 = \frac{1}{\langle N \rangle} + \frac{1}{\Omega^2} \int \int \omega(\theta) d\Omega_1 d\Omega_2 \qquad (1)$$

We used another, more straightforward method, to determine the variance: we counted the number of objects in boxes with the desired area and directly measured the variance from them. We have weighted the results from each box by its 'useful' surface. The resulting rms for each of our five fields are listed as errors in table 2. These empirical values of the rms are bigger than $\sqrt{N}$: on average 1.55 times for brighter objects and 1.5 for fainter. Yee et al. 1986 found that for $r < 22.0$, rms $= 1.35\sqrt{N}$.

The sum of the number of objects found within the central regions of the five fields give us 91 objects against $59.7 \pm 12.5$ with $19.5 < R < 22$ (i.e. an empirical excess of $2.5\sigma$) and 215 objects against $220.4 \pm 22.3$ with $22 < R < 23.75$. Therefore, no significant excess is found at fainter magnitudes. In order to determine the statistical significance of the excess found at brighter magnitudes, we have constructed an empirical probability distribution law. This distribution is formed by the sum of the objects contained in all the possible combinations of five boxes, each of them from a different field. We have normalized the results of each box by the useful surface of the radiogalaxy box of that field. Then we have excluded all the combinations of boxes that have an added useful surface smaller than a given threshold which is less or equal than the sum of the useful surfaces of the radiogalaxy boxes (94% of 5 2 arcmin × 2 arcmin boxes). This exclusion must be made carefully: if we set a very low exclusion threshold the tails of the distribution become dominated by the boxes with small area, which after being surface-corrected, have their rms scaled as $\sim N$ when in fact it should scale as $\sim \sqrt{N}$. On the other hand, if we set the threshold too high we exclude too many boxes and the results are not representative of the full image. After trying with several values of the exclusion threshold, we see that the excess is significant, in the most conservative case, on a $\gtrsim 99.5\%$ level for the 2 arcmin scale. The shape of the distribution so obtained is very close to a Gaussian: from the 2.5$\sigma$ found above we would expect a 99.4% probability assuming Gaussian statistics which is very close to the found value.

## 6. Comparison with other results

The environment of QSOs and BL Lac objects at ($0.9 < z < 1.5$) has been studied by several authors (Hintzen et al. 1991, Boyle & Couch 1993, Hutchings et al. 1993, Fried et al. 1993). Our work is, as far as we know, the first one studying the environment of radiogalaxies at similar redshifts. It is therefore very interesting to compare our results with those obtained for QSOs and BL Lac objects.

Hintzen et al. (1991) looked for clustering around 16 radio-loud QSOs with $0.9 < z < 1.5$ and found 32 objects with $R < 23$ within 15 arcsec where the expectation was 19.3. We, on the other hand, found 6 objects in our four fields where we would expect 6.0. No excess is present.

Boyle & Couch (1993) have searched for associations between faint galaxies and 27 radio-quiet QSOs with $0.9 < z < 1.5$. No significant excess is found, in agreement with results for lower redshifts (Ellingson et al. 1991). Adding the results from all their fields, they found 2687 objects with $R < 23$ instead of 2721.3 within a radius of 120 arcsec around the QSOs. With the same constraints, we found in our five fields 506 objects against 439.4. Even assuming that the distribution of galaxies is not poissonian (see sec. 5.) and that it has a rms equal to $\approx 1.5\sqrt{N}$, this is a 2.1$\sigma$ excess ( more than half of this excess is within the range $19.5 < R < 22$: 212 objects against 173.8). The environment of radio-quiet QSOs seems thus to be quite different from that of radiogalaxies. Given that these radio-quiet QSOs are intrinsically fainter than the radio-loud QSOs of Hintzen et al., Boyle & Couch proposed the observation of bright radio-quiet QSOs in order to find out whether the environment correlates with optical luminosity or radio power. In a certain sense, we have performed that test; the objects studied by Boyle and Couch are on average $\gtrsim 1$ mag intrinsically more luminous than our radiogalaxies, what makes unlikely that the different environments of radio-quiet and radio-loud QSOs correlate with optical luminosity.

Fried et al. (1993) did not detect any significant excess of $R < 22$ galaxies around five BL Lac objects with $\langle z \rangle = 0.97$. This seems not to be the case for our sample: we found 73 objects instead of 44.7 with $19.5 < R < 22$ (a $2.7\sigma$ empirical excess), radius $< 60$ arcsec. There is no excess at fainter magnitudes: within the same angular scale and for $22 < R < 23.75$ there are 158 objects instead of 154. If we followed the reasoning of Fried et al. — that is, that due to the high redshift of the objects and the bright limiting magnitude, any excess found would be due to gravitational lensing — then we would have found a clear proof of our radiogalaxies being gravitationally lensed objects.

## 7. Conclusions

We have taken deep $R$-band images of fields around five $R \approx 21.4$ radiogalaxies: 0956+47, 1217+36, 3C256, 3C324 and 3C294. 0956+47 is found to show a double nucleus and we confirm (McCarthy 1991) that it probably forms part of a group of galaxies at z=1.026. We give another measure of the magnitude for 1217+36 ($R = 21.6$) and suggest the revision of its classification as a radiogalaxy. We found a statistically significant excess of bright galaxies around our sample of radiogalaxies (see Sec. 5.). We found 91 objects with $19.5 < R < 22$ within a box of 2 arcmin around the central radiogalaxies of our five fields; the expectation based on the average density should be 59.7. We determined empirically the significance level and it corresponds to $\gtrsim 99.5\%$. It is remarkable that no excess is detected at fainter magnitudes: for $22 < R < 23.75$ we only found 215 objects against 220.4 within a 2 arcmin box. This strongly suggests that the excess is not physically associated to the galaxies but due to intervening groups and then related to gravitational lensing.

NB, EMG and JIGS acknowledge financial support from the Spanish DGICYT, project PB92-0741. Partial financial support for this project was provided by the Comission of the European Union and their Human Capital and Mobility Contract CHRX-CT92-0033. NB acknowledges a Spanish M.E.C. Ph.D. scholarship. LC acknowledges a M.E.C.-Fullbright postdoctoral fellowship. The Isaac Newton Telescope is operated by the Royal Greenwich Observatory at the Spanish Observatorio del Roque de los Muchachos of the Instituto de Astrofísica de Canarias, on behalf of the Science and Engineering Research council of the United Kingdom and the Netherlands Organization for Scientific Research.

FIGURE CAPTIONS

Figure 1. Number counts versus magnitude from the average of the sum of the five fields. The errors are the rms of the sum.

Figure 2. Optical $R$-band image of 0956+47 showing its double nucleus. Axis units are in pixels, one pixel being equal to 0.57 arcsec. North is up and east is to the left. The contour levels are spaced $1\sigma_{sky}$, starting at $2\sigma_{sky}$ over the mean sky value. $\sigma_{sky}$ is the rms of the sky brightness and corresponds to 26.6 mag arcsec$^{-2}$.

Figure 3. Optical $R$-band images of the fields of a) 1217+36 (512,618) , b) 0956+47 (519,623), c) 3C256 (502,624), d) 3C324 (524,603), and e) 3C294 (500,611) showing an area $4 \times 4$ arcmin$^2$ centered on each radiogalaxy. Each axis subdivision is equal to 100 pixels ($\approx 57$ arcsec). The orientation and the contour levels are the same of the previous image.

TABLE 1. General data for each field

| Radiogalaxy | z | R | S(%) | $N_{19.5<R<22}$ | $N_{22<R<23.75}$ |
|---|---|---|---|---|---|
| 1217+36 | 1.2[a] | 21.56 | 83.4 | 288 | 1248 |
| 0956+47 | 1.026 | 21.22 | 87.2 | 341 | 1226 |
| 3C256 | 1.819 | 21.51 | 77.5 | 316 | 1246 |
| 3C324 | 1.206 | 21.45 | 75.3 | 317 | 996 |
| 3C294 | 1.78 | —[b] | 80.9 | 261 | 883 |

[a]Estimated from the $K-z$ relation (see references in text)
[b]Photometry is not possible because of a nearby saturated star

Notes to Table 1.

Redshifts (taken from references in text), $R$-band magnitudes, S(%) is the percentage of useful surface from a $1200 \times 1090$ frame. $N$ is the number of objects in the whole frame.

TABLE 2. Data for $2 \times 2$ arcmin$^2$

| Field | $N_b$ | $N_f$ | s(%) | $n_b$ | $ne_b$ | $n_f$ | $ne_f$ |
|---|---|---|---|---|---|---|---|
| 1217+36 | $11.7 \pm 4.5$ | $50.5 \pm 8.8$ | 100 | 25 | $11.7 \pm 4.5$ | 55 | $50.5 \pm 8.8$ |
| 0956+47 | $13.2 \pm 5.7$ | $47.4 \pm 9.5$ | 100 | 21 | $13.2 \pm 5.7$ | 54 | $47.4 \pm 9.5$ |
| 3C256 | $13.8 \pm 7.8$ | $54.2 \pm 12.1$ | 94 | 21 | $12.8 \pm 7.6$ | 50 | $50.8 \pm 11.7$ |
| 3C324 | $14.2 \pm 4.9$ | $44.6 \pm 9.3$ | 87 | 14 | $12.3 \pm 4.6$ | 32 | $38.8 \pm 8.6$ |
| 3C294 | $10.9 \pm 4.9$ | $36.8 \pm 11.5$ | 89 | 10 | $9.7 \pm 4.6$ | 24 | $32.9 \pm 10.9$ |

Notes to Table 2.

$N_b$ and $N_f$ are the average of bright ($19.5 < R < 22$) and faint ($22 < R < 23.75$) objects within a box (the errors are the empirically found rms). s(%) is the percentage of useful surface of the box centered on the radiogalaxy. $n_b$,$n_f$ are the number of bright and faint objects found and $ne_b$,$ne_f$ the expectation within that box.